\documentclass[twocolumn]{aastex631}
\usepackage[T1]{fontenc}
\usepackage{graphicx}
\usepackage[]{natbib}
\usepackage{amsmath, multirow}
\usepackage{hyperref}
\usepackage[colorinlistoftodos]{todonotes}

\newcommand{\hst}{\textit{HST}}
\newcommand{\jwst}{\textit{JWST}}
\newcommand{\Spitzer}{\textit{Spitzer}}

\newcommand{\db}{\texttt{DENSE BASIS}}

\newcommand{\galfit}{\texttt{Galfit}}
\newcommand{\grizli}{\texttt{Grizli}}
\newcommand{\astrodrizzle}{\texttt{Astrodrizzle}}
\newcommand{\cgs}{erg\,s$^{-1}$\,cm$^{-2}$}

\newcommand{\um}{$\mu$m}
\newcommand{\macs}{MACS~J1149.5$+$2223}
\newcommand{\jd}{MACS1149-JD1}
\newcommand{\lya}{Lyman-$\alpha$}

\newcommand{\Niii}{N~{\sc iii}]~$\lambda$1747,1749}
\newcommand{\Oiiia}{[O~{\sc iii}]~$\lambda$4959}
\newcommand{\Oiiib}{[O~{\sc iii}]~$\lambda$5007}

\def\otf{[\ion{O}{3}] 88$~\mu\mbox{m}$}

\begin{document}
\title{Star Formation at the Epoch of Reionization with CANUCS: The ages of stellar populations in {\jd}}
\shorttitle{CANUCS: Star Formation in {\jd}}
\shortauthors{Brada\v{c} et al.}

\author[0000-0001-5984-0395]{Maru\v{s}a Brada{\v c}}
\affiliation{University of Ljubljana, Department of Mathematics and Physics, Jadranska ulica 19, SI-1000 Ljubljana, Slovenia}
\affiliation{Department of Physics and Astronomy, University of California Davis, 1 Shields Avenue, Davis, CA 95616, USA}

\author[0000-0002-6338-7295]{Victoria Strait}
\affiliation{Cosmic Dawn Center (DAWN), Denmark}
\affiliation{Niels Bohr Institute, University of Copenhagen, Jagtvej 128, DK-2200 Copenhagen N, Denmark}

\author[0000-0002-8530-9765]{Lamiya Mowla}
\affiliation{Whitin Observatory, Department of Physics and Astronomy, Wellesley College, 106 Central Street, Wellesley, MA 02481, USA}
\affiliation{Dunlap Institute for Astronomy and Astrophysics, 50 St. George Street, Toronto, Ontario M5S 3H4, Canada}

\author[0000-0001-9298-3523]{Kartheik G. Iyer}
\affiliation{Dunlap Institute for Astronomy and Astrophysics, 50 St. George Street, Toronto, Ontario M5S 3H4, Canada}

\author{Gaël Noirot}
\affiliation{Department of Astronomy and Physics and Institute for Computational Astrophysics, Saint Mary's University, 923 Robie Street, Halifax, Nova Scotia B3H 3C3, Canada}

\author[0000-0002-4201-7367]{Chris Willott}
\affiliation{NRC Herzberg, 5071 West Saanich Rd, Victoria, BC V9E 2E7, Canada}

\author[0000-0003-2680-005X]{Gabe Brammer}
\affiliation{Cosmic Dawn Center (DAWN), Denmark}
\affiliation{Niels Bohr Institute, University of Copenhagen, Jagtvej 128, DK-2200 Copenhagen N, Denmark}

\author[0000-0002-4542-921X]{Roberto Abraham}
\affiliation{David A. Dunlap Department of Astronomy and Astrophysics, University of Toronto, 50 St. George Street, Toronto, Ontario, M5S 3H4, Canada}

\author[0000-0003-3983-5438]{Yoshihisa Asada}
\affiliation{Department of Astronomy and Physics and Institute for Computational Astrophysics, Saint Mary's University, 923 Robie Street, Halifax, Nova Scotia B3H 3C3, Canada}
\affiliation{Department of Astronomy, Kyoto University, Sakyo-ku, Kyoto 606-8502, Japan}

\author[0000-0001-8325-1742]{Guillaume Desprez}
\affiliation{Department of Astronomy and Physics and Institute for Computational Astrophysics, Saint Mary's University, 923 Robie Street, Halifax, Nova Scotia B3H 3C3, Canada}

\author{Vince Estrada-Carpenter}
\affiliation{Department of Astronomy and Physics and Institute for Computational Astrophysics, Saint Mary's University, 923 Robie Street, Halifax, Nova Scotia B3H 3C3, Canada}

\author[0000-0001-9414-6382]{Anishya Harshan}
\affiliation{University of Ljubljana, Department of Mathematics and Physics, Jadranska ulica 19, SI-1000 Ljubljana, Slovenia}

\author[0000-0003-3243-9969]{Nicholas S. Martis}
\affiliation{Department of Astronomy and Physics and Institute for Computational Astrophysics, Saint Mary's University, 923 Robie Street, Halifax, Nova Scotia B3H 3C3, Canada}
\affiliation{National Research Council of Canada, Herzberg Astronomy \& Astrophysics Research Centre, 5071 West Saanich Road, Victoria, BC, V9E 2E7, Canada}

\author[0000-0002-7547-3385]{Jasleen Matharu}
\affiliation{Cosmic Dawn Center (DAWN), Denmark}
\affiliation{Niels Bohr Institute, University of Copenhagen, Jagtvej 128, DK-2200 Copenhagen N, Denmark}

\author{Adam Muzzin}
\affiliation{Department of Physics and Astronomy, York University, 4700 Keele St. Toronto, Ontario, M3J 1P3, Canada}

\author[0009-0009-4388-898X]{Gregor Rihtar\v{s}i\v{c}}
\affiliation{University of Ljubljana, Department of Mathematics and Physics, Jadranska ulica 19, SI-1000 Ljubljana, Slovenia}

\author[0000-0001-8830-2166]{Ghassan T. E. Sarrouh}
\affiliation{Department of Physics and Astronomy, York University, 4700 Keele St. Toronto, Ontario, M3J 1P3, Canada}

\author[0000-0002-7712-7857]{Marcin Sawicki}
\affiliation{Department of Astronomy and Physics and Institute for Computational Astrophysics, Saint Mary's University, 923 Robie Street, Halifax, Nova Scotia B3H 3C3, Canada}

\begin{abstract}

We present measurements of stellar populations properties of a $z=9.1$ gravitationally lensed galaxy {\jd} using deep {\it JWST} NIRISS slitless spectroscopy as well as NIRISS and NIRCam imaging from the CAnadian NIRISS Unbiased Cluster Survey (CANUCS). The galaxy is split into four components. Three  magnified ($\mu\sim 17$) star-forming components are unresolved, giving intrinsic sizes of $<50\mbox{pc}$. In addition, the underlying extended  component contains the bulk of the stellar mass, formed the majority of its stars $\sim 50 \mbox{Myr}$ earlier than the other three components and is not the site of most active star formation currently. The NIRISS and NIRCam resolved photometry does not confirm a strong Balmer break previously seen in {\Spitzer}. The NIRISS grism spectrum has been extracted for the entire galaxy and shows a clear continuum and Lyman-break, with no {\lya} detected. 

\end{abstract}


\section{Introduction}
\label{sec:intro}

Tracing star formation to the earliest times has been a long-standing goal of extragalactic astronomy. In particular, studying the onset of
star formation is of importance not only for galaxy formation models
but also for studies of the early
universe. {\Spitzer} and the Hubble Space Telescope ({\hst}) played
a unique role in determining the onset of star formation of galaxies at redshift $z\gtrsim 6$ (e.g., \citealp{bradac20} for a review). 

The James Webb Space Telescope ({\jwst}, \citealp{gardner23}) is revolutionizing studies of the early onset of star formation in high-redshift galaxies. With the expanded sensitivity, filter set and wavelength coverage compared to \Spitzer, \jwst\ can trace the full spectral energy distribution, and in some cases distinguish strong line emission from breaks due to evolved stars (e.g., \citealp{laporte23}). Early results from \jwst\ appear to show a higher than expected ultraviolet luminosity density at $z>10$ \citep{harikane23a, donnan23}.  There have also been claims for the presence at $7<z<9$ of massive galaxies with strong Balmer breaks in the CEERS survey \citep{labbe23, lovell23, boylankolchin23}. However, studies from other \jwst\ surveys with comparable or larger volumes such as JADES, EPOCHS and CANUCS do not find such a high density of massive galaxies with strong Balmer breaks at these redshifts (\citealp{endsley23b, trussler23}; \citealp[in prep.]{desprez23}).

One of the most intriguing objects showing a potential Balmer break from previous  \hst\ and \Spitzer\ studies is the
$z=9.1$ galaxy {\jd} behind the cluster {\macs}.  {\jd} was originally discovered in {\hst} and shallow
{\Spitzer} data in \citet{zheng12}.  It was later detected in both channel 1 and channel 2
{\Spitzer} bands using deeper data
\citep{surfsup,huang16,zheng17,hoag18} and its
redshift was spectroscopically measured with the {\otf} line using ALMA by \citet{hashimoto18}. With early data, it was concluded that the nebular emission lines
are redshifted out of both {\Spitzer} bands (at $z>9$), yet the galaxy showed a
strong color excess. It was
therefore highly likely that old ($\sim 300\mbox{Myr}$) stellar populations are causing the
red rest-frame optical colors
(\citealp{hashimoto18,hoag18,huang16}). This was surprising, given the galaxy would need to start forming a significant amount of stars
shortly after the Big
Bang ($\sim 250\mbox{Myr}$). In addition, the cold dust content of the galaxy
was constrained to be modest from observations taken with ALMA, making
dust an unlikely cause of red {\Spitzer} color \citep{hashimoto18}.

{\jd} was recently observed as part of the CAnadian NIRISS Unbiased Cluster Survey (CANUCS, \citealp{willott22}) with the NIRCam and NIRISS instruments onboard {\jwst}. The data provides superior photometry compared to what was possible with {\Spitzer}. In addition, NIRISS spectra with its coverage from 1 to $2.5$ {\um} allow us to investigate the rest frame UV spectrum, including searching for the presence of potential {\lya} line (previously mentioned in \citealp{hashimoto18}). Here we describe these data and analysis of the stellar properties of {\jd}. 

The paper is structured as follows. In Section~\ref{sec:data} we
present the data used in this paper and in 
Section~\ref{sec:dataanalysis} we describe the analysis of the photometric and spectroscopic data. In
Section~\ref{sec:results} we present the main science results.  We summarize in
Section~\ref{sec:conclusions} and give photometry and SED fitting results in the tables in Appendix~\ref{sec:app}.

Throughout the paper we assume a
$\Lambda$CDM cosmology from \citet{planck18} with $\Omega_{\rm m}=0.310$ and Hubble constant
$H_0=67.7{\rm\ kms^{-1}\:\mbox{Mpc}^{-1}}$.

\section{Data}
\label{sec:data}
 {\jwst} NIRISS and NIRCam observations of {\jd} were taken from 10th-22nd May 2023 as part of the NIRISS GTO Program \#1208, The Canadian NIRISS Unbiased Cluster Survey (CANUCS, \citealt{willott22},  \dataset[DOI]{https://doi.org/10.17909/ph4n-6n76}). 
 
 The field was observed with NIRCam imaging using filters F090W, F115W, F150W, F200W, F277W, F356W, F410M, and F444W with exposure times of $6.4 \mbox{ks}$ each, reaching S/N between 5 and 10 for a $m_{\rm AB} = 29$ point source. We also utilized archival data of {\tt HST} imaging from Hubble Frontier Fields \citep{lotz17}, GLASS \citep{treu15}, and SN Refsdal \citep{kelly15} follow-up observations (HST-GO-13504, PI Lotz; HST-GO-13790, PI Rodney;  HST-GO-13459, PI Treu; HST GO-14041, PI Kelly).

 To reduce the imaging data we use the photometric pipeline that will be presented in  Brammer
et al. (in prep.), which also provides a compilation of
the {\jwst} ERO photometric data released to date. Briefly, the
raw data  has been reduced using the
public Grism redshift \& line analysis software {\grizli} \citep{grizli23,grizli23b}, which
masks imaging artifacts, provides astrometric calibrations
based on the Gaia Data Release 3 catalog, and shifts images
using {\astrodrizzle}. The method closely follows the one outlined in \citet{valentino23}. We show the cutouts of {\jd} in Fig.~\ref{fig:cutouts}. 
 
 Observations also consist of two NIRISS pointings, one centred on the cluster centre containing {\jd} and the other coincident with a flanking field. Each pointing is observed with the GR150R and GR150C grisms through the F115W, F150W and F200W filters. Exposure times for the cluster field  are  19240 seconds in each of the three filters. We also process all the NIRISS imaging and slitless spectroscopy with \texttt{Grizli}. \texttt{Grizli} performs full end-to-end processing of space-based slitless spectroscopic datasets. For full details see e.g., \citet{matharu21,noirot23, matharu23}. In summary, raw data is downloaded from the Mikulski Archive for Space Telescopes (MAST) and pre-processed for cosmic rays, flat-fielding, sky subtraction, astrometric corrections and alignment. Contamination models (which correct for overlapping spectra from nearby sources) for each pointing are then generated and subtracted for each grism spectrum of interest. From these images we extract the spectrum of {\jd}.

For the magnification estimate, we use the model from \citet{finney18}, which uses Hubble Frontier field data (referred to {\tt Bradacv4}). The corresponding magnification is $\mu_{\rm best}=17^{+3}_{-5}$ ($68\%$ confidence). It is in agreement with the
median magnification $\mu_{\rm models}=13^{+140}_{-7}$ from the six other publicly available
post-HFF lens models ({\tt CATSv4.1, Sharonv4cor, Keetonv4, Williamsv4, Diegov4.1, GLAFICv3}) published on MAST\footnote{\url{https://archive.stsci.edu/prepds/frontier/lensmodels/}}. Throughout the paper we correct for gravitational lensing magnification $\mu_{\rm best}$ all properties as appropriate unless otherwise noted. 

\section{Data Analysis}
\label{sec:dataanalysis}
\subsection{Photometry and SED Fitting}
The photometry is derived using the updated
zero-points, and corrected for Milky Way extinction. We use F150W and 
F200W NIRISS filters  as well as F150W,
F200W, F277W, F356W, F410M, and F444W NIRCam filters (in other {\jwst} filters {\jd} is not/barely detected) and {\hst} upper limits for the entire source.  Since the object is resolved into three distinct clumps and a smooth galaxy component, we perform a photometric fit using {\galfit} \citep{peng10}. We forward model the source assuming four components (three point sources for the clumps and a sersic profile for the diffused light), convolve them with the PSF and determine their parameters to minimise the residuals. We measure empirical PSF determined from the stars.  

\begin{figure}
      \includegraphics[width=0.5\textwidth]{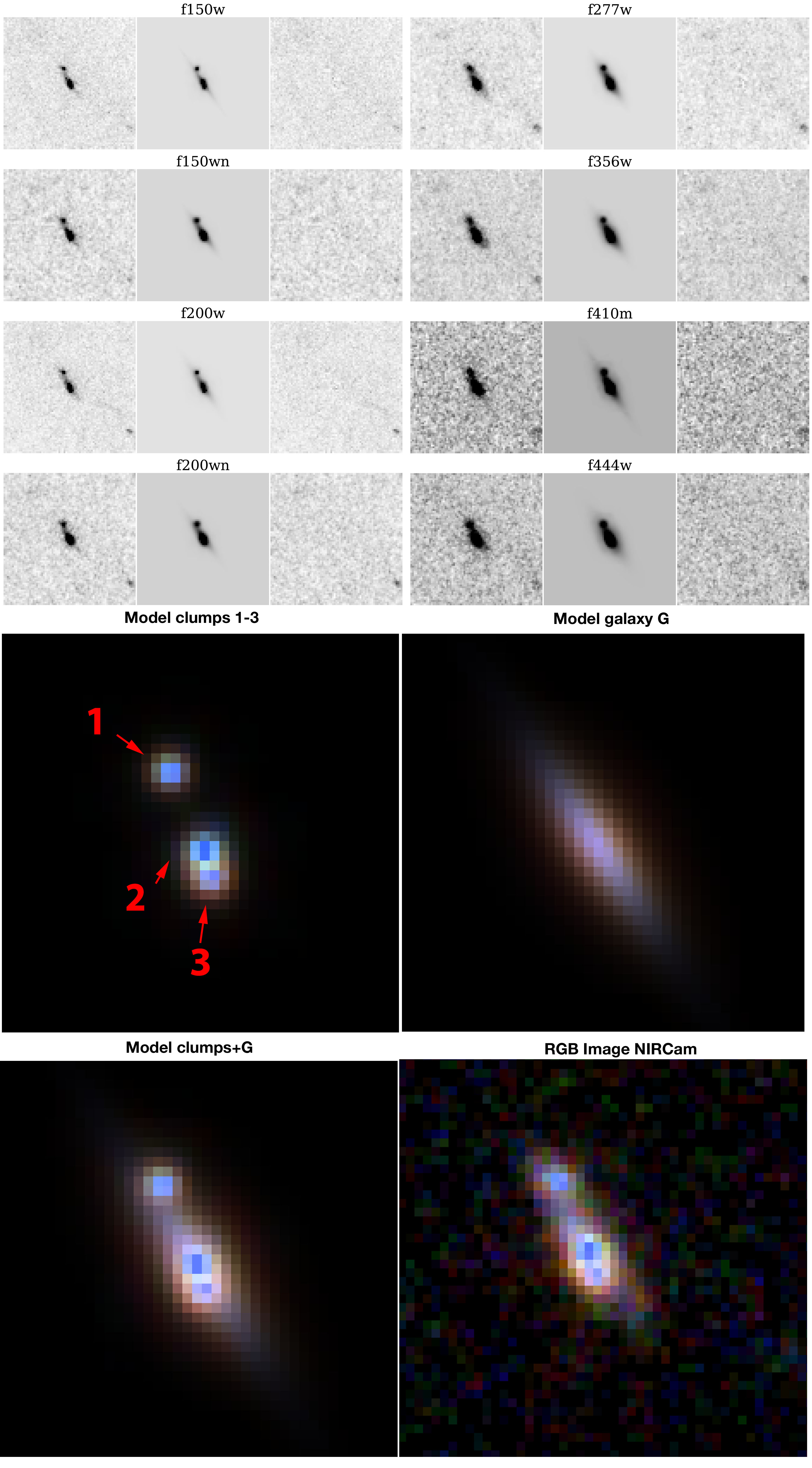}
  \caption{Images of the  three clumps and underlying galaxy component in {\jd}. Shown are different filters, {\galfit} models and residuals (top panels), RGB (always using F150W, F277W, and F444W filters) model of the clumps and the galaxy (middle panels) and RGB model and RGB NIRCam image (bottom panels). The galaxy consists of at least 3 clumps (all marked in the middle panel), all of which are unresolved and an underlying smooth component.  An upper limit to the magnified size (FWHM) of the clumps is $0.05\arcsec$. The intrinsic (demagnified) size upper limit is $<50\mbox{pc}$.}
    \label{fig:cutouts}
 \end{figure}

The resulting models and residuals are shown in Fig.~\ref{fig:cutouts}. Residuals from the fits are negligible, confirming the original visual
impression that the three compact sources are unresolved and an additional smooth component is present. The agreement between NIRISS and NIRCam fluxes in the two overlapping filters is a confirmation of the robustness of photometry. Resolved photometry is necessary, as global spectral energy distribution (SED) fitting can bias stellar masses when young stellar population outshine the first episodes of star formation (e.g., \citealp{sorba18, gimenez-arteaga23, narayanan23}).
Photometric properties are given in Table~\ref{tab:phot}.   

\begin{figure}
\centerline{\includegraphics[width=0.5\textwidth]{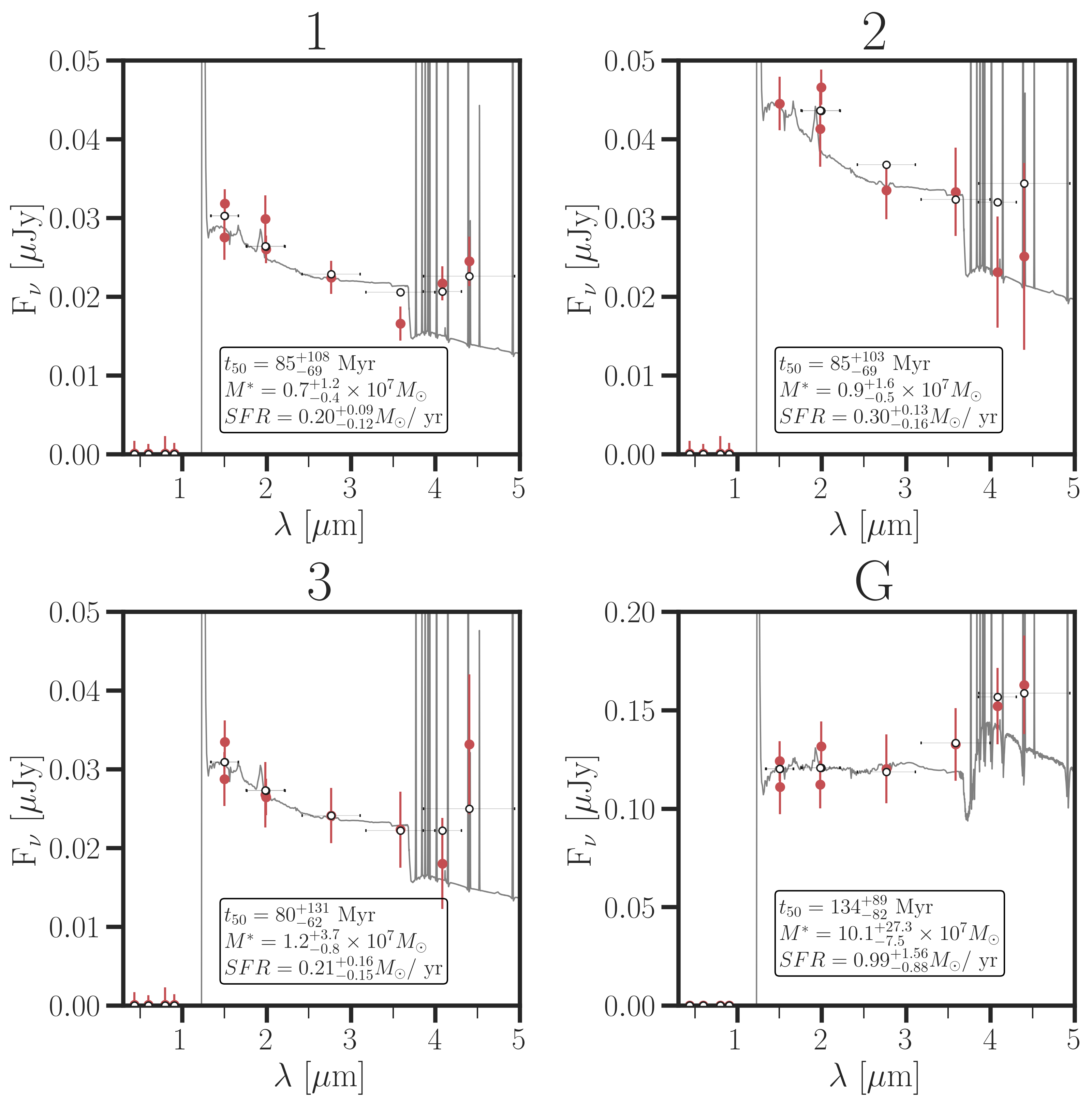}}
 \caption{Results of the SED fitting for the three clumps (labelled 1-3) and the smooth light component (G). Shown are measured fluxes (i.e., we do not correct them for magnification) for both NIRCam and NIRISS imaging in red and SED predicted fluxes in open circles in units of $\mu\mbox{Jy}$. Derived stellar properties are given in the inset.}
    \label{fig:SEDfit}
\end{figure}

SEDs derived from our photometry were analysed using the {\db}
method \citep{iyer17, iyer19} to determine
nonparametric star formation histories (SFHs), masses, and ages for our 
sources in {\jd}. We adopt the
Calzetti attenuation law \citep{calzetti01} and a Chabrier IMF \citep{chabrier03}. We fix the redshift to that found by the {\otf} line in \citet{hashimoto18}, $9.1096 \pm 0.0006$. All other parameters are left free. The primary
advantage of using {\db} with nonparametric SFHs is that they allow us
to account for flexible stellar populations. Both photometry and SED fit are shown in Fig.~\ref{fig:SEDfit}.

\subsection{Grism spectroscopy}
\begin{figure*}
\centerline{\includegraphics[width=1.0\textwidth]{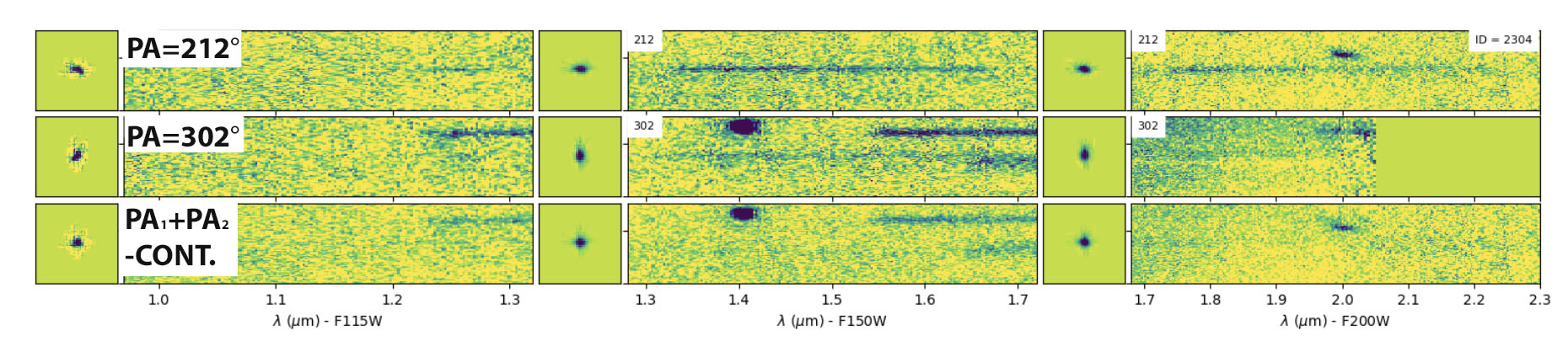}}
      \centerline{\includegraphics[width=1.0\textwidth]{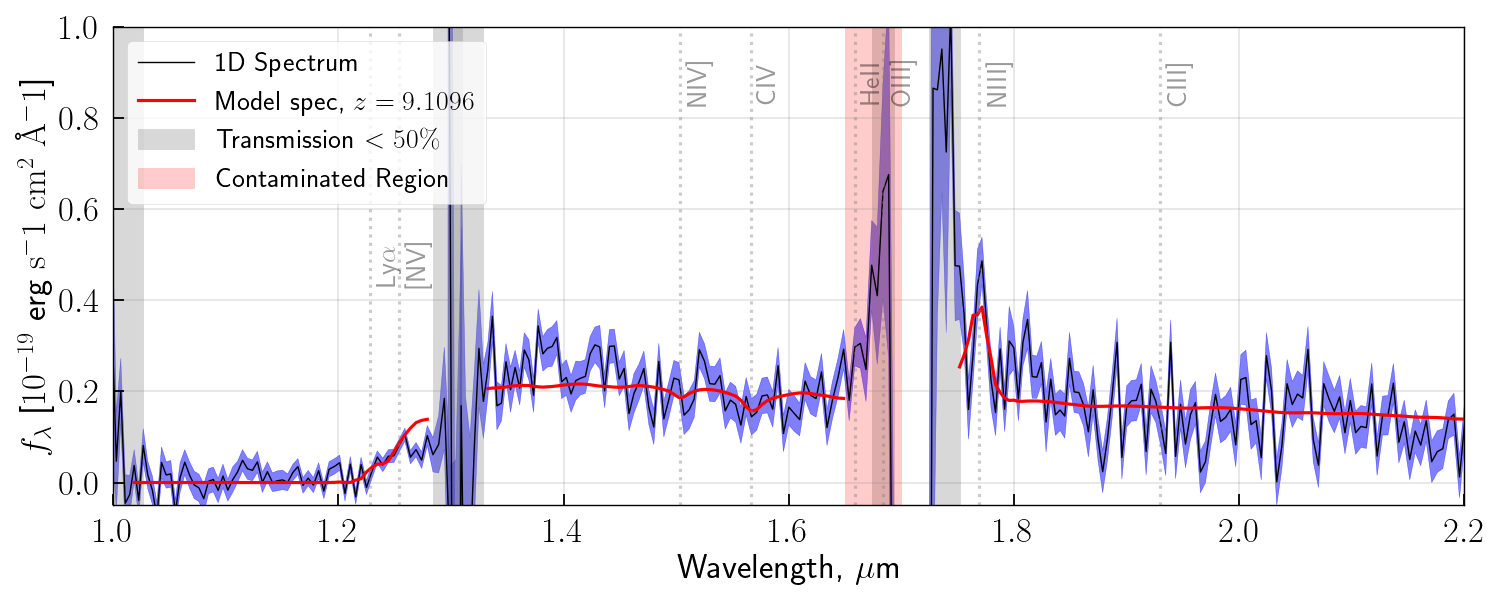}}
  \caption{NIRISS grism spectrum of {\jd}. {\bf Top: } 2D spectrum of {\jd} is shown in two orientations ($\mbox{PA}=212\deg$ top and $302\deg$ middle) and three filters. Direct images are also shown. The bottom row shows a combined spectrum with continuum emission from the object subtracted. All images are contamination subtracted, the residual contamination is from objects below the imaging detection threshold and objects outside the FoV of the direct imaging. {\bf Bottom: } 1D spectrum extracted (black line with uncertainties in blue) and modelled given fixed redshift (red line). The positions of potential lines are marked. Only {\Niii} line is possibly detected in $\mbox{PA}=212\deg$; the other orientation is contaminated and the spectrum falls on the edge of the detector (see top). The region where contamination subtraction failed is marked in red and the region between half power wavelengths at which the transmission in each filter falls below 50\% of its peak value is marked in grey.}
    \label{fig:GRISMSpec}
 \end{figure*}

To extract the NIRISS spectrum of the source we also use the {\grizli} package. 
The {\grizli} reduction steps of the NIRISS data includes the creation
of a NIRISS direct image mosaic from which diffraction spikes
of bright sources are masked. Following \citet{noirot23} source detection is performed
on the NIRISS mosaic image with the {\tt Source Extractor} \citep{Bertin:1996ww} python wrapper {\tt sep} \citep{barbary16}, using the
default detection parameters implemented in {\grizli} (a detection
threshold {\tt `threshold'} of $1.8\sigma$ above the global background RMS,
a minimum source area in pixels {\tt ‘minarea’} of 9, and deblending
parameters {\tt 'deblend\_cont'} and {\tt ‘deblend\_ntresh’} of 0.001 and 32, respectively). Matched aperture photometry on the available NIRISS filters is performed at
the same stage. From this NIRISS imaging catalogue, the position of
sources that contaminate the spectrum of {\jd} are used to locate spectral traces in the grism data.  The
spectral continua of the sources are modelled using an iterative polynomial
fitting of the data for contamination estimate and removal.
The 2D and extracted 1D {\jd} spectra with contamination removal and modelled spectrum are shown in Fig.~\ref{fig:GRISMSpec}.

\begin{figure}
\centerline{\includegraphics[width=0.5\textwidth]{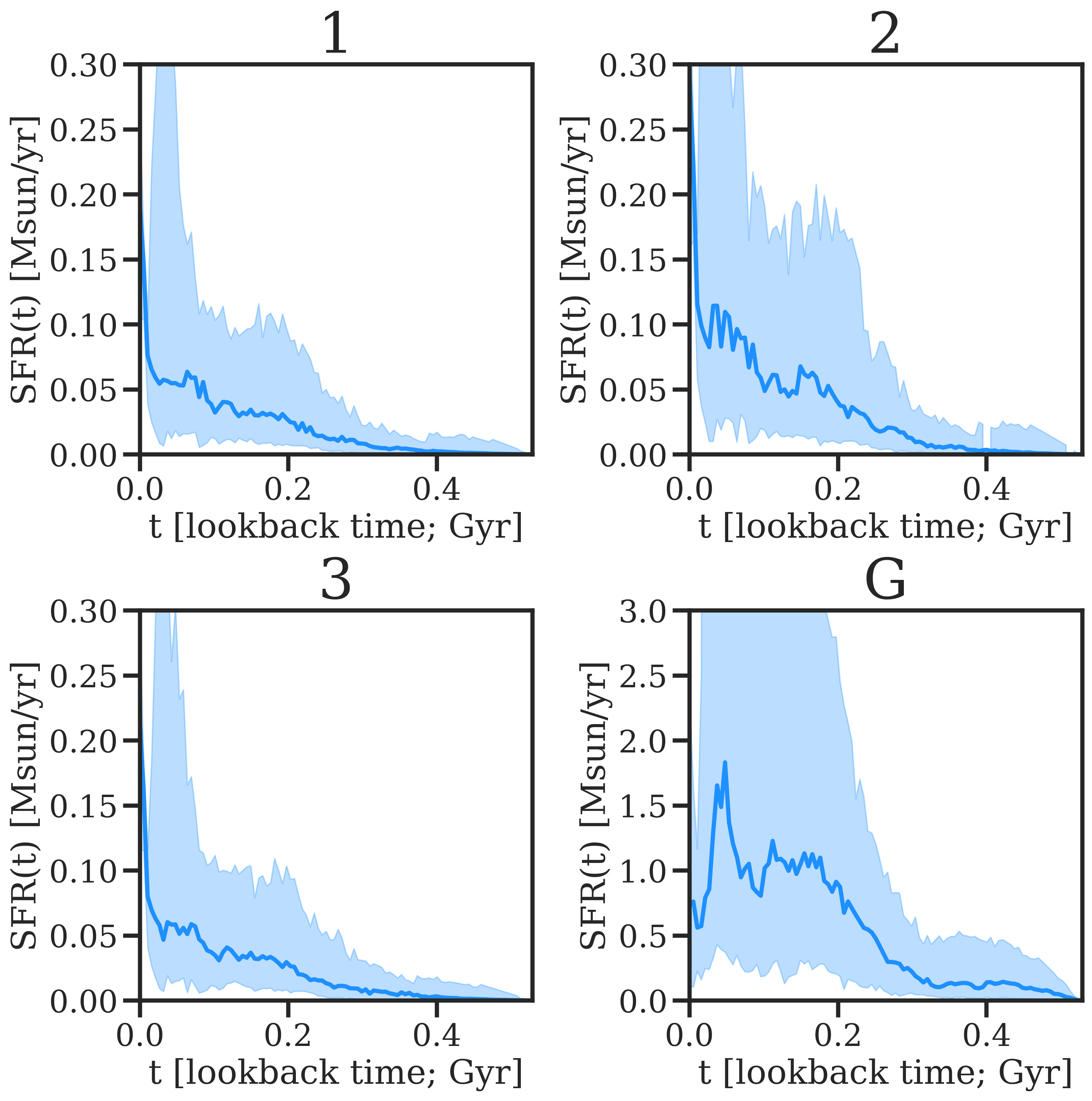}}
 \caption{Star formation histories for the four components. While the three star-forming clumps have similar star-formation histories, the underlying galaxy component is different.}
    \label{fig:sfrh}
\end{figure}

\begin{figure}
\centerline{\includegraphics[width=0.5\textwidth]{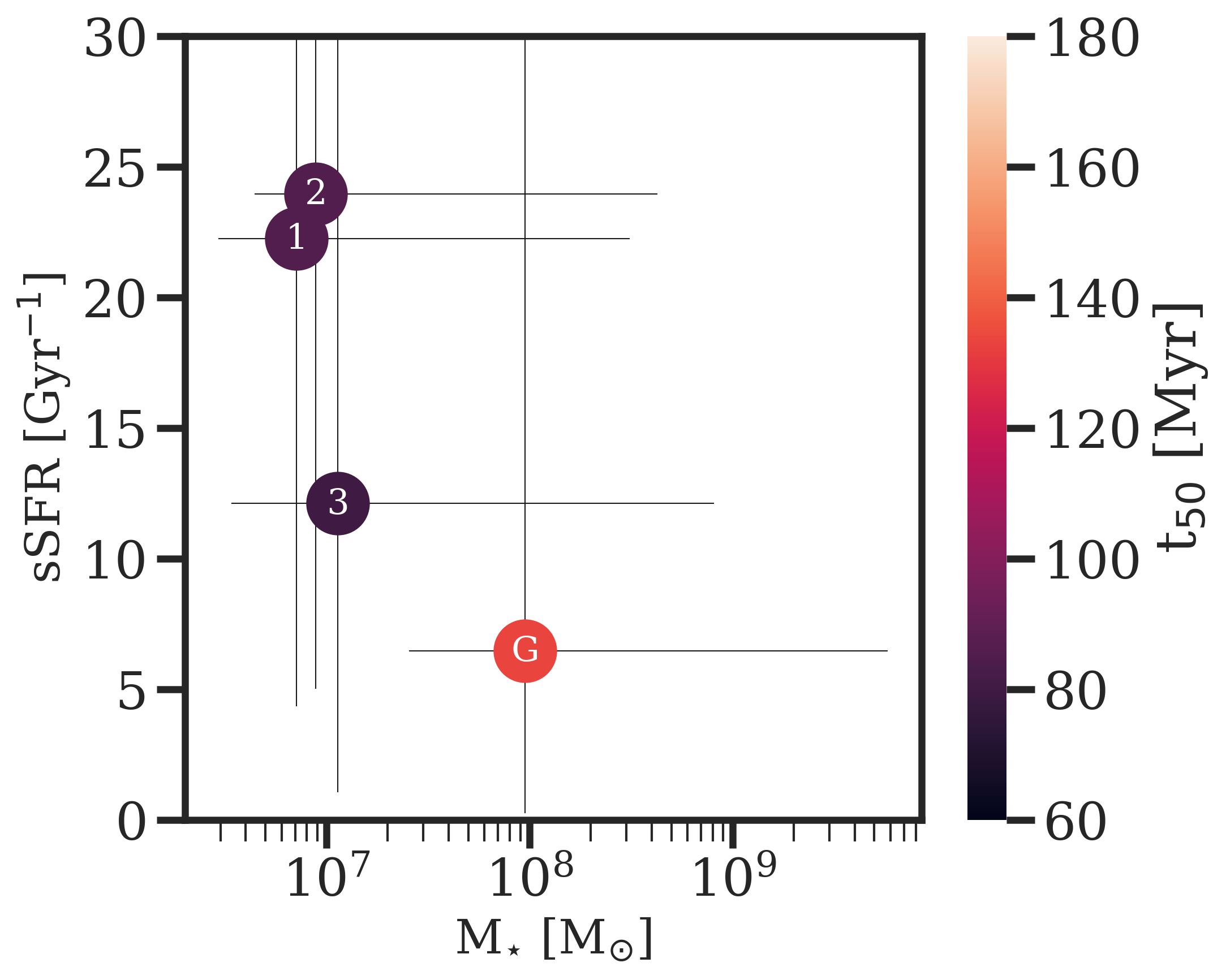}}
 \caption{Specific star formation rate (sSFR) vs. stellar mass ($M_*$) plot for the four components. All three unresolved components show  similar stellar ages, while the underlying galaxy component shows an older stellar population  (albeit with large errorbars, see also Table~\ref{tab:prop}).}
    \label{fig:sSFRage}
\end{figure}

\begin{figure}
\centerline{\includegraphics[width=0.5\textwidth]{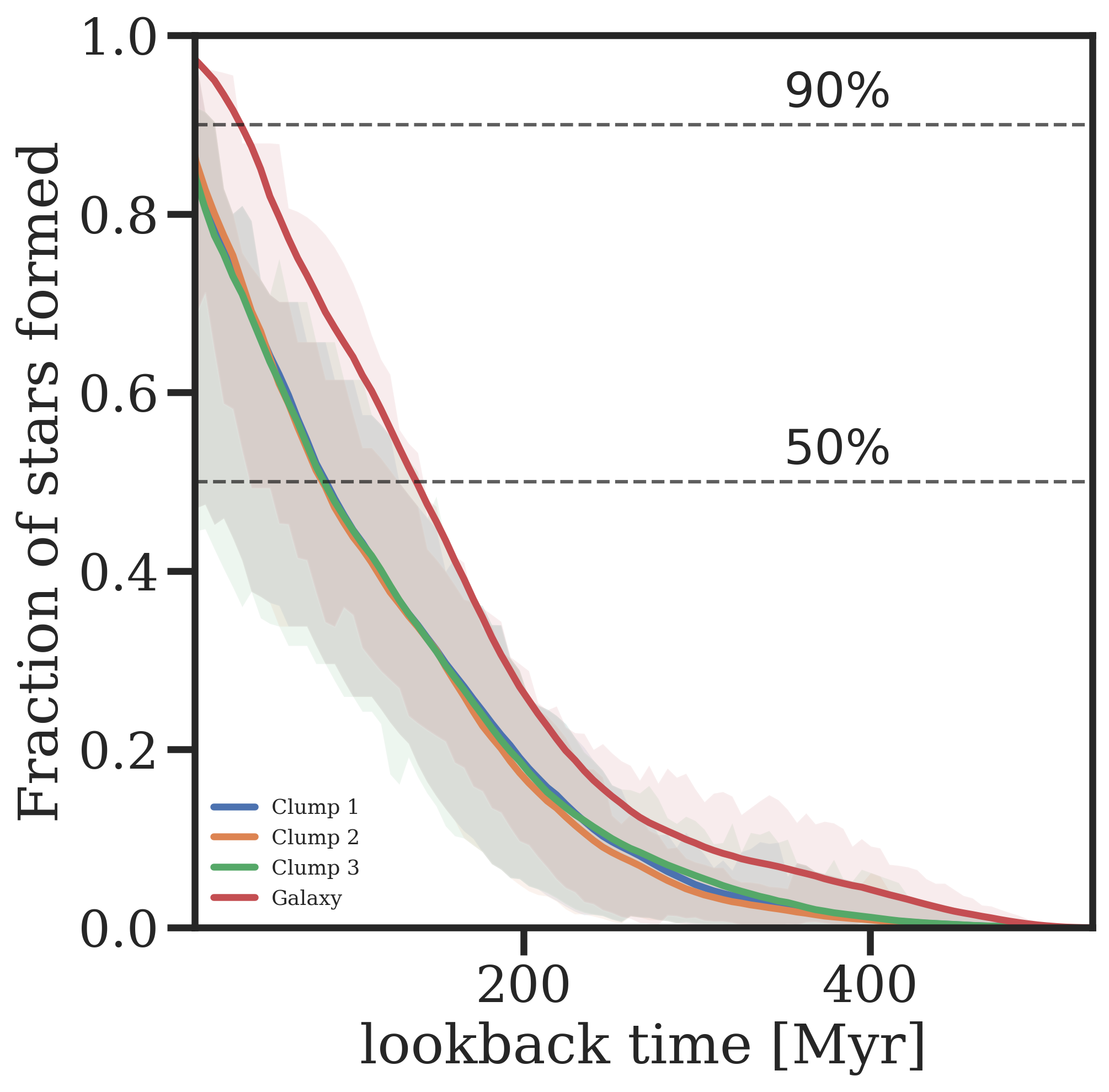}}
 \caption{Mass fraction of stars formed as a function of look-back time for all four components (1-blue, 2-orange, 3-green, galaxy-red). In \citealp{hashimoto18} the authors conclude that the bulk of the stellar mass was produced within a short period
corresponding to the redshift interval $12 < z < 16$, with a dominant stellar component that
formed at the look-back time of $\sim 290\mbox{Myr}$. These new measurements show somewhat younger ages, with the oldest component (G) forming 50\% of its total mass at $t_{50}=134^{+ 89}_{- 82}$.}
    \label{fig:fraction}
\end{figure}
\section{Results}
\label{sec:results}

\subsection{Spatially Resolved Star Formation History}
\label{sec:resphot}

Using our photometry (Table \ref{tab:phot}) we now determine the stellar properties of each individual component. By determining $\beta_{\rm UV}$ slopes based on NIRCam F150W, F200W, and F227W fluxes, we see that the three clumps have different properties than the underlying galaxy component. The three clumps have $\beta_{\rm UV,phot}$ measured between $-2.5$ and, $-2.8$, whereas the galaxy itself is redder with  $\beta_{\rm UV,phot}=-1.9\pm 0.2$ (Table~\ref{tab:prop}). The values are consistent with  with other observations from {\jwst} (e.g., \citealp{topping23,endsley23,bouwens23b,franco23}). 

Using {\db} we also perform the SED fit and determine
nonparametric star formation histories. All four components have intrinsic (corrected for magnification) stellar masses between $~5\times 10^6$ and $10^8M_{\odot}$ and star formation rates (SFR) between $0.2-1M_{\odot}\mbox{yr}^{-1}$. While the errorbars are large and there is still a possibility that all components have the same star formation histories, there is nevertheless a hint that the galaxy itself (G) has started to form the bulk of the stars earlier (Fig.~\ref{fig:sfrh}). This component also has the highest stellar mass (Fig.~\ref{fig:sSFRage}). In Fig.~\ref{fig:fraction} we also show the mass fraction of stars formed as a function of lookback time. Once again, the error-bars are large, but there is an indication that the underlying galaxy has formed the bulk of its stellar masses earlier than the clumps, which are still actively star-forming.

The galaxy formed 50\% of its total mass at $t_{50}=134^{+ 89}_{- 82} \mbox{Myr}$.  In \citet{hashimoto18}, the bulk of the stellar population was determined to have formed at a look-back time of $\sim 250\mbox{Myr}$. The main reason is that the relative flux measured red-ward of $\sim 4000\mbox{\AA}$ has decreased, making the potential Balmer-break less pronounced. We measure the Balmer break of galaxy G (based on fluxes in F444W and F277W, the latter being mostly emission line free) of $\Delta mag_{\rm AB} = 0.3\pm0.2\mbox{mag}$ ($F_\nu({\rm F444W})/F_\nu({\rm F277W}) = 1.4\pm0.2$). This is lower compared to {\Spitzer} measurements from \citet{kokorev22} of $\Delta mag_{\rm AB} = 0.5 \pm 0.2$, from \citet{zheng17} (also used in \citealp{hashimoto18}) $\Delta mag_{\rm AB}  >1.3$ (1-$\sigma$) , from ASTRODEEP \citep{dicriscienzo17} $\Delta mag_{\rm AB}  >0.7$ (3-$\sigma$) and from \citep{huang16} $\Delta mag_{\rm AB} = 0.8\pm 0.4$. All {\Spitzer} measurements are the average of the older G component and all the clumps, though the former dominates the flux. This discrepancy is unlikely caused by emission lines, as both NIRCam F444W and {\Spitzer} Channel 2 have similar throughputs at the red-end, hence entering {\Oiiia} emission line (where both instruments have a throughput of 20\%) could not play a role (for {\Oiiib} both are similarly at 1\%). We think the most likely source of discrepancy is the contamination modelling which in the case of {\Spitzer}'s large PSF is difficult. 

\subsection{Grism Spectrum}

We clearly detect the continuum and Lyman-break in the NIRISS spectrum at the expected redshift. In Fig.~\ref{fig:GRISMSpec} we show 1D spectral extraction with a fitted model at the redshift determined by \citet{hashimoto18}. However, even if we let the redshift be determined by the NIRISS data alone, we still recover the same redshift ($z=9.2\pm 0.1$).  We have also searched for the {\lya} emission line that was indicated in \citet{hashimoto18} at $12,267.4\mbox{\AA}$ with an integrated (lensed) flux of $4.3\pm 1.1 \times 10^{-18}$\cgs. We do not detect any lines at that wavelength. From the sensitivity of our observations, such a line would have been detected. We do, however, detect a line in one orientation $\mbox{PA}=212\deg$ at $17,700\mbox{\AA}$  which corresponds to {\Niii},  with the flux of $4.6\pm 0.6 \times 10^{-18}$\cgs. Unfortunately, the other orientation is contaminated and furthermore, the spectrum is located towards the edge of the detector. Hence, we consider this line tentative. 

The combined UV beta slope measured from the NIRISS spectrum between rest-frame wavelengths of $1400-1600\mbox{\AA}$ and $1800-2000\mbox{\AA}$ (we assume a similar spectral range as used for photometry, excluding the part of the spectrum in the detector gap)  is $\beta_{\rm UV,spec}=-2.3\pm 0.5$. This is consistent with the average photometric measurements done for individual clumps (Table.~\ref{tab:prop}).

The spectrum also shows a softening of the Lyman-break in the vicinity of {\lya}, very likely caused by a largely neutral IGM \citep{mason20, curtislake23, heintz23}. Unfortunately, the break falls at the gap between the two filters, hence we cannot characterize it fully.

%

\section{Conclusions} \label{sec:conclusions}
The gravitationally lensed galaxy {\jd} at $z=9.1096 \pm 0.0006$ has been well studied in the past. {\Spitzer} data were showing what seemed to be a strong Balmer break, meaning that the dominant stellar component formed about 290\mbox{Myr} earlier (or around $240\mbox{Myr}$ after the Big Bang, \citealp{surfsup, huang16, hoag18, hashimoto18}). 

New {\jwst} observations with NIRISS and NIRCam reveal that the galaxy consists of three unresolved (with intrinsic sizes $<50 \mbox{pc}$) star-forming clumps and an underlying extended galaxy component. We individually perform SED fitting of all four components (Fig.~\ref{fig:SEDfit}, Table~\ref{tab:prop}). The galaxy component (G) is showing somewhat older stellar population, albeit with large errobars. This component (i) contains the bulk of the stellar mass, (ii) likely formed the majority of its stars $\sim 50 \mbox{Myr}$ earlier than the other components and (iii) is not the site of the most recent star formation. 

NIRISS spectrum of {\jd} shows a clear detection of the continuum and Lyman-break. However, we do not detect the {\lya} line previously reported in \citet{hashimoto18}. Given that NIRISS spectra have low spectral resolution {\lya} could still be present, though at a lower flux than previously reported. 

In conclusion, {\jd} is a highly magnified, intrinsically faint galaxy at $z=9.1$. It shows properties that are consistent with other galaxies detected with {\jwst} (e.g., \citealp{bunker23b,topping23}); however, its true nature was only revealed through resolved SED fitting.  While strong Balmer breaks can be present at high redshift, they are rare \citep{laporte23,looser23,strait23,endsley23b}. With the newest data for {\jd}, a strong Balmer break is excluded. 

\section*{Acknowledgements}
MB, GR, and AH acknowledge support from the ERC Grant FIRSTLIGHT and Slovenian national research agency ARRS through grants N1-0238 and P1-0188. MB  acknowledges support from the program HST-GO-16667, provided through a grant from the STScI under NASA contract NAS5-26555.
This research was enabled by grant 18JWST-GTO1 from the Canadian Space Agency and funding from the Natural Sciences and Engineering Research Council of Canada.
 This research used the Canadian Advanced Network For Astronomy Research (CANFAR) operated in partnership by the Canadian Astronomy Data Centre and The Digital Research Alliance of Canada with support from the National Research Council of Canada the Canadian Space Agency, CANARIE and the Canadian Foundation for Innovation. The Cosmic Dawn Center (DAWN) is funded by the Danish National Research Foundation under grant No. 140. This work utilizes gravitational lensing models produced by PIs Brada\v{c}, Natarajan \& Kneib (CATS), Merten, Zitrin, Sharon, Williams, Keeton, Bernstein and Diego, and the GLAFIC group. This lens modeling was partially funded by the HST Frontier Fields program conducted by STScI. STScI is operated by the Association of Universities for Research in Astronomy, Inc. under NASA contract NAS 5-26555. The lens models were obtained from the Mikulski Archive for Space Telescopes (MAST).
\section*{Data Availability}

 The data is available at DOI: \url{10.17909/ph4n-6n76}

\facilities{HST (ACS,WFC3), JWST (NIRCam, NIRISS)}

\bibliographystyle{aasjournal}
\bibliography{bibliogr_cv,bibliogr_highz}

\appendix

\section{Photometry and SED fitting}
\label{sec:app}
In Table~\ref{tab:phot} we list photometry and in Table~\ref{tab:prop} derived quantities and results of SED fitting of {\jd}. All the procedures are described in the main text. 

\begin{deluxetable}{lcccc} 

\tabletypesize{\footnotesize}
\tablecolumns{5}
\tablewidth{0pt}
\tablecaption{{\jwst} photometry (fluxes for all four components) of {\jd}.\label{tab:phot}}
\tablehead{\colhead{Filter} & \colhead{C \#1}& \colhead{C \#2} & \colhead{C \#3}& \colhead{G}  \vspace{-0.2cm} \\
\colhead{} & \colhead{(nJy)}& \colhead{(nJy)} & \colhead{(nJy)}& \colhead{(nJy)}}
\startdata
\multicolumn{5}{c}{NIRISS}\\
F150W & $28 \pm 3$ & $45 \pm 3$ & $29 \pm 3$ & $124 \pm 11$ \\
F200W & $30 \pm 3$ & $41 \pm 5$ & $27 \pm 4$ & $112 \pm 10$ \\
\hline
\multicolumn{5}{c}{NIRCam}\\
F150W & $32 \pm 2$ & $54 \pm 3$ & $33 \pm 3$ & $111 \pm 16$ \\
F200W & $26 \pm 2$ & $47 \pm 2$ & $26 \pm 3$ & $132 \pm 12$ \\
F277W & $22 \pm 2$ & $33 \pm 3$ & $24 \pm 2$ & $120 \pm 8$ \\
F356W & $17 \pm 2$ & $33 \pm 4$ & $22 \pm 3$ & $133 \pm 12$ \\
F410M & $22 \pm 2$ & $23 \pm 6$ & $18 \pm 7$ & $152 \pm 20$ \\
F444W & $24 \pm 3$ & $25 \pm 10$ & $33 \pm 8$ & $163 \pm 13$ \\
\hline 
\enddata
\end{deluxetable}

\begin{deluxetable*}{lcccc} 
\tabletypesize{\footnotesize}
\tablecolumns{5}
\tablewidth{0pt}
\tablecaption{Global properties of {\jd} and stellar properties of individual clumps\label{tab:prop}.}
\tablehead{\colhead{Parameter} & \colhead{}& \colhead{}& \colhead{}& \colhead{} \\ 
\colhead{} &  \colhead{C \#1}& \colhead{C \#2} & \colhead{C \#3}& \colhead{G}}
\startdata
R.A. (deg) & & & &177.3899418  \\
Decl. (deg) & & & & 22.4126885 \\
$z_{\rm spec}$\tablenotemark{\small a} & & &  & $9.1096 \pm 0.0006$ \\
$\mu_{\rm best}$\tablenotemark{\small b} & & &  & $17^{+3}_{-5}$ \\
\hline
$\beta_{\rm UV,phot}$ & $-2.6 \pm 0.1$ & $-2.8 \pm 0.1$ & $-2.5 \pm 0.2$ & $-1.9 \pm 0.2$ \\
$M^* (M_{\odot})$ & $0.7^{+1.1}_{-0.4} \times 10^7 $& $0.9^{+1.6}_{- 0.4} \times 10^7$ & $1.1^{+3.6}_{- 0.8} \times 10^7$ & $9.5^{+25}_{- 7.0} \times 10^7$\\
$SFR (M_{\odot}/\mathrm{~yr})$ & $0.20^{+ 0.09}_{- 0.12}$ & $0.30^{+ 0.13}_{- 0.15}$ & $0.21^{+ 0.16}_{- 0.15}$ & $0.99^{+ 1.45}_{- 0.88}$\\
$t_{50} (\mathrm{~Myr})$ & $85^{+ 108}_{- 69}$ & $85^{+ 103}_{- 69}$ & $80^{+ 131}_{- 62}$ & $134^{+ 89}_{- 82}$
\enddata
\tablenotetext{a}{From \citet{hashimoto18}.}
\tablenotetext{b}{Model {\tt Bradacv4} \citet{finney18}.}
\end{deluxetable*}

\end{document}